\documentclass[12pt]{article}

\usepackage[margin=1.0in]{geometry}

\usepackage[nosort]{cite}

\usepackage{amssymb,latexsym}

\usepackage{epstopdf}

\usepackage{graphics,epsfig}

\usepackage{color}
\usepackage{xcolor}

\usepackage{amsmath,amsfonts}

\usepackage{mathrsfs}

\usepackage{slashed}

\DeclareMathAlphabet{\mathpzc}{OT1}{pzc}{m}{it}

\usepackage{feynmp}
\usepackage{feynmp-auto}

\let\a=\alpha \let\b=\beta \let\g=\gamma \let\d=\delta \let\e=\epsilon
\let\z=\zeta  \let\th=\theta  \let\k=\kappa
\let\l=\lambda \let\m=\mu \let\n=\nu \let\x=\xi \let\p=\pi 
\let\s=\sigma   \let\f=\phi  
 
        \let\Th=\Theta 
\let\X=\Xi  \let\S=\Sigma  \let\Y=\Psi
 
\let\la=\label  
  
\def\nn{\nonumber} \def\bd{\begin{document}} \def\ed{\end{document}}
\def\ds{\documentstyle} \let\fr=\frac \let\bl=\bigl \let\br=\bigr
\let\Br=\Bigr \let\Bl=\Bigl
\let\bm=\bibitem
\let\na=\nabla
\def\tU{{\widetilde U}}
\let\pa=\partial \let\ov=\overline
\def\ie{{\it i.e.\ }}
\newcommand{\be}{\begin{equation}}
\newcommand{\ee}{\end{equation}}
\def\ba{\begin{array}}
\def\ea{\end{array}}
\def\ft#1#2{{\textstyle{{\scriptstyle #1}\over {\scriptstyle #2}}}}
\def\fft#1#2{{#1 \over #2}}
\def\F#1#2{{ F_{#1}^{(#2)} }}
\def\cF#1#2{{ {\cal F}_{#1}^{(#2)} }}

\def\R{{\bf R}}
\def\sst#1{{\scriptscriptstyle #1}}
\def\oneone{\rlap 1\mkern4mu{\rm l}}
\def\e7{E_{7(+7)}}
\def\td{\tilde}
\def\wtd{\widetilde}
\def\im{{\rm i}}
\def\bog{Bogomol'nyi\ }
\newcommand{\ho}[1]{$\, ^{#1}$}
\newcommand{\hoch}[1]{$\, ^{#1}$}
\newcommand{\bea}{\begin{eqnarray}}
\newcommand{\eea}{\end{eqnarray}}
\newcommand{\ra}{\rightarrow}
\newcommand{\lra}{\longrightarrow}
\newcommand{\Lra}{\Leftrightarrow}
\newcommand{\ap}{\alpha^\prime}
\newcommand{\bp}{\tilde \beta^\prime}
\newcommand{\cB}{{\cal B}}
\newcommand{\cO}{{\cal O}}
\newcommand{\vecx}{\vec{x}}
\newcommand{\vecy}{\vec{y}}
\newcommand{\vecp}{\vec{p}}
\newcommand{\vecq}{\vec{q}}
\newcommand{\tr}{{\rm tr} }
\newcommand{\Tr}{{\rm Tr} }
\newcommand{\NP}{Nucl. Phys. }

\newcommand{\cL}{{\cal L}}
\newcommand{\cA}{{\cal A}}
\newcommand{\cT}{{\cal T}}
\newcommand{\cR}{{\cal R}}
\newcommand{\cD}{{\cal D}}
\newcommand{\cH}{{\cal H}}

\def\Cb{\bar{C}}

\def\sst#1{{\scriptscriptstyle #1}}
\def\0{{\sst{(0)}}}
\def\1{{\sst{(1)}}}
\def\2{{\sst{(2)}}}
\def\3{{\sst{(3)}}}
\def\4{{\sst{(4)}}}
\def\5{{\sst{(5)}}}
\def\6{{\sst{(6)}}}
\def\7{{\sst{(7)}}}
\def\8{{\sst{(8)}}}
\def\9{{\sst{(9)}}}
\def\p{{\sst{(p)}}}
\def\q{{\sst{(q)}}}
\def\ve{\varepsilon}
\def\vf{\varphi}
\def\F{\Phi}
\def\wg{\wedge}

\def\e{\epsilon}
\def\barl{\bar{l}}

\def \bi{\bibitem}
\def \la {\label}

\def \l {\lambda}
\def\foot{\footnote}
\def \tl  {{\tilde \l}}
\def \sql {{\sqrt \l}}
\def \adss {$AdS_5 \times S^5$\ }
\newcommand{\rf}[1]{(\ref{#1})}
\def \ov {\over}

\def\th{\theta}
\def\Th{\Theta}
\def\vth{\vartheta}
\def\btheta{{\bar\theta}}
\def\ttheta{{{\tilde\theta}}}
\def\bttheta{{{\bar\ttheta}}}
\def\vth{\vartheta}

\def\ra{\rightarrow}
\def\N{\nabla}
\def\F{{\cal F}}
\def\uM{\underline{M}}
\def\uA{\underline{A}}
\def\uN{\underline{N}}
\def\uP{\underline{P}}
\def\ua{\underline{a}}
\def\ub{\underline{b}}
\def\uc{\underline{c}}
\def\ud{\underline{d}}
\def\ue{\underline{e}}
\def\uf{\underline{f}}
\def\ui{\underline{i}}
\def\uj{\underline{j}}
\def\uk{\underline{k}}
\def\ul{\underline{l}}
\def\ual{\underline{\alpha}}
\def\ube{\underline{\beta}}
\def\um{\underline{m}}
\def\un{\underline{n}}
\def\up{\underline{p}}
\def\uq{\underline{q}}
\def\ur{\underline{r}}
\def\us{\underline{s}}
\def\umu{\underline{\mu}}
\def\unu{\underline{\nu}}
\def\ula{\underline{\l}}
\def\uka{\underline{\k}}
\def\usi{\underline{\s}}
\def\urh{\underline{\r}}
\def\cc{\circ}
\def\eqv{\equiv}

\def\ni{\noindent}

\def\Ep{E^{{}^{(+)}}}
\def\Em{E^{{}^{(-)}}}

\def\Mp{M^{{}^{(+)}}}
\def\Mm{M^{{}^{(-)}}}

\def \ha{{1\ov 2}}

\def\r{\rho}

\def\Y{{\rm Y}}
\def\X{{\rm X}}
\def\tY{\tilde{\rm Y}}
\def\tX{\tilde{\rm X}}
\def\dY{\dot{\rm Y}}
\def\dX{\dot{\rm X}}

\def \J {\mathcal{J}}
\def \del {\partial}

\def\dF{\dot{F}}
\def\dG{\dot{G}}
\def\df{\dot{f}}
\def \E {{\cal E}}
\def \S {{\cal S}}
\def \J {{\cal J}}

\def\ms{\mathcal{S}}
\def\mj{\mathcal{J}}
\def\soj{\fr{\ms}{\mj}}
\def \R {{\bf R}}
\def \om {\omega}
\def \bE {\bar E}
\def \x {{\cal X}}

\def \bi{\bibitem}
\def \la {\label}

\def \l {\lambda}
\def\foot{\footnote}
\def \tl  {{\tilde \l}}
\def \sql {{\sqrt \l}}
\def \adss {$AdS_5 \times S^5$\ }
\def \ov {\over}

\def \varpi {{\rm w}}

\def\thb{\bar{\theta}}
\def\Thb{\bar{\Theta}}
\def\barp{\bar{p}}
\def\barq{\bar{q}}
\def\barc{\bar{c}}
\def\bard{\bar{d}}
\def\bare{\bar{e}}

\def\thb{\bar{\theta}}
\def\Thb{\bar{\Theta}}
\def\mb{\bar{\m}}
\def\ab{\bar{\a}}
\def\zb{\bar{z}}
\def\psib{\bar{\psi}}
\def\barl{\bar{l}}
\def\barp{\bar{p}}
\def\barq{\bar{q}}
\def\barc{\bar{c}}
\def\bard{\bar{d}}
\def\baru{\bar{u}}

\def\e{\epsilon}
\def\wb{\bar{w}}
\def\lb{\bar{\l}}
\def\Jb{\bar{J}}
\def\Nb{\bar{N}}
\def\Zb{\bar{Z}}
\def\pab{\bar{\pa}}

\def\At{\tilde{A}}
\def\Bt{\tilde{B}}
\def\Ct{\tilde{C}}
\def\Dt{\tilde{D}}
\def\Et{\tilde{E}}
\def\Ft{\tilde{F}}
\def\Gt{\tilde{G}}
\def\Ht{\tilde{H}}
\def\Kt{\tilde{K}}
\def\Mt{\tilde{M}}
\def\Nt{\tilde{N}}
\def\Rt{\tilde{R}}
\def\at{\tilde{a}}
\def\bt{\tilde{b}}
\def\ct{\tilde{c}}
\def\dt{\tilde{d}}
\def\et{\tilde{e}}
\def\ft{\tilde{f}}
\def \ztt{\tilde{\z}}
\def \zetat{\tilde{\zeta}}
\def\htil{\tilde{h}}
\def\gt{\tilde{g}}
\def\nt{\tilde{n}}
\def\mut{\tilde{\mu}}
\def\nut{\tilde{\nu}}
\def\pht{\tilde{\f}}
\def\Phit{\tilde{\Phi}}
\def\vft{\tilde{\vf}}

\def\rht{\tilde{\rho}}

\def\asth{\hat{*}}
\def\phh{\hat{\phi}}

\def\bA{{\bf A}}

\def\ola{\overleftarrow}
\def\ora{\overrightarrow}
\def\alt{\tilde{\a}}

\def\eh{\hat{e}}
\def\eph{\hat{\e}}
\def\ph{\hat{p}}
\def\alh{\hat{\a}}
\def\beh{\hat{\b}}
\def\gah{\hat{\g}}
\def\Fh{\hat{F}}
\def\muh{\hat{\m}}
\def\nuh{\hat{\n}}
\def\thh{\hat{\th}}
\def\rhh{\hat{\r}}
\def\dh{\hat{d}}
\def\ih{\hat{i}}
\def\jh{\hat{j}}
\def\hh{\hat{h}}
\def\nh{\hat{n}}
\def\gh{\hat{g}}
\def\kh{\hat{k}}
\def\deh{\hat{\d}}
\def\wh{\hat{w}}
\def\lah{\hat{\l}}
\def\Ah{\hat{A}}
\def\Gh{\hat{G}}
\def\Kh{\hat{K}}
\def\Nh{\hat{N}}
\def\Rh{\hat{R}}
\def\Ch{\hat{C}}
\def\Omh{\hat{\Omega}}

\def\xh{\hat{x}}

\def\ps{\rlap{\, /}\;\,p }
\def\ks{\rlap{\, /}\;\,k }

\def\gym{g_{YM}}

\def\adot{\dot{a}}
\def\bdot{\dot{b}}
\def\bpa{\bar{\pa}}

\def\pr{\prime}
\def\ssk{\medskip}
\def\clb{\color{blue}}
\def\clr{\color{red}}
\def\clg{\color{green}}
\def\clp{\color{purple}}
\def\clc{\color{cyan}}
\def\clm{\color{magenta}}
\def\cly{\color{yellow}}

\def\bfA{{\bf A}}
\def\bfB{{\bf B}}
\def\bfK{{\bf K}}
\def\bfU{{\bf U}}
\def\bfX{{\bf X}}
\def\bfY{{\bf Y}}
\def\bfZ{{\bf Z}}
\def\bfg{{\bf g}}
\def\bfn{{\bf n}}

\def\bfL{{\bf L}}
\def\bfJ{{\bf J}}
\def\bfE{{\bf E}}
\def\bfF{{\bf F}}
\def\bfQ{{\bf Q}}
\def\bfC{{\bf C}}
\def\bfeps{{\bf \epsilon}}

\def\bsk{\bigskip}
\def\ssk{\medskip}

\def\Ec{{\cal E}}

\begin{document}

\overfullrule=0pt
\parskip=2pt
\parindent=12pt
\headheight=0in \headsep=0in \topmargin=0in
\oddsidemargin=0in

\vspace{ -3cm}
\thispagestyle{empty}

 \vspace{0.1cm}

\setcounter{equation}{0}
\setcounter{footnote}{0}
\setcounter{section}{0}

\begin{center}

{\Large\bf   Black hole entropy from non-Dirichlet sectors, and a bounce solution}

\vskip 0.8cm

%
%
I. Y. Park
\\

\vspace{0.3cm}

\vspace{0.3cm}
{\it Department of Applied Mathematics,
Philander Smith College 
                               \\
Little Rock, AR 72202, USA \\
inyongpark05@gmail.com
}

 \vspace{.5cm}

\end{center}

 \vspace{0.1cm}

\begin{abstract}

 The relevance of gravitational boundary degrees of freedom and their dynamics in gravity quantization and black hole information has been explored in a series of recent works. In this work we further progress by  focusing keenly on the genuine gravitational boundary degrees of freedom as the origin of black hole entropy. Wald's entropy formula is scrutinized, and the reason that Wald's formula correctly captures the entropy of a black hole examined. Afterwards, limitations of Wald's method are discussed; a coherent view of entropy based on boundary dynamics is presented. The discrepancy observed in the literature between holographic and Wald's entropies is addressed. We generalize the entropy definition so as to handle a time-dependent black hole. Large gauge symmetry plays a pivotal role. Non-Dirichlet boundary conditions and gravitational analogues of Coleman-De Luccia bounce solutions are central in identifying the microstates and differentiating the origins of entropies associated with different classes of solutions. The result in the present work leads to a view that black hole entropy is entanglement entropy in a thermodynamic setup.

\end{abstract}
\newpage





\section{Introduction}

The difficulty in cracking the black hole information (BHI) paradox \cite{Hawking:1974sw}\cite{Hawking:1976ra} (see, e.g., \cite{Buoninfante:2021ijy} for a recent review on the paradox) is attributed to several factors. One may attribute this difficulty, among other things, to the intricacies in quantizing gravity, given that quantized gravity must be a proper framework for tackling the paradox (and some other long-standing problems). Another factor is the enigma of identifying the black hole microstates that are responsible for the black hole entropy (see, e.g., \cite{Solodukhin:2011gn} for a review). Still another reason can be found in that the boundary dynamics was incompletely dealt with in the past; a more recently explored solution of the BHI puzzle pivots on the boundary dynamics \cite{Hatefi:2012bp,Park:2017wiw,Park:2018xtt,Park:2019lbj}. These considerations are in fact related, as will be discussed in more detail. The boundary configuration induces the corresponding gravitational analogue of a Coleman-De Luccia (CdL) bounce solution \cite{Coleman:1980aw}.\footnote{The same name,``bounce," is used for the ``Lorentzian bounce solutions" in the gravitational context \cite{Ambrus:2005nm,Haggard:2014rza,Christodoulou:2016vny,Bianchi:2018mml,BenAchour:2020mgu,BenAchour:2020gon,Malafarina:2017csn}. The main conceptual difference of our view is the instontonic physics aspect of a CdL bounce solution.} Non-perturbative bounce physics as well as boundary dynamics is indispensable for a proper understanding of BHI.

The present work is not directly about gravity quantization. It is nevertheless the issues encountered while quantizing gravity that provide crucial threads to follow in tracking not only the quantization itself but also several long-standing puzzles such as BHI paradox. To illustrate, one way of connecting different moving parts is as follows. As well known, gravity has a large amount of gauge symmetry, diffeomorphism: large gauge symmetry as well as regular one. They both play important roles in quantization, with the former crucial for the present entropy analysis. The large gauge symmetry is also behind a curtain of several recently analyzed phenomenon. For instance, it provides one of the key rationales for considering non-Dirichlet boundary conditions associated with non-trivial boundary dynamics \cite{Park:2018xtt}. 
Conventionally, a Dirichlet boundary condition has been used in gravitational theories and is a crucial component in defining an isolated system. Once one includes non-Dirichlet conditions, there is a host of issues including the crucial task of determining the physical sector of the theory, a proper handling of which is an essential part of quantization. For the present work we focus on the attributes of the large gauge symmetry and their implications for the black hole microstates.

We anticipate that the large gauge symmetry and boundary degrees of freedom should be responsible for the black hole microstates and their associated entropy. What makes this identification possible is the fact that the boundary degrees of freedom are part of the gravitational degrees of freedom. This view was explained in earlier works of \cite{Hatefi:2012bp,Park:2017wiw,Park:2018xtt,Park:2019lbj} (and is in spirit of \cite{Park:1999xz}): a given background of the bulk theory is reduced by the physical state condition (see \cite{Higuchi:1991tk} for a linear-level analysis of a dS background), and the reduced action describes the fluctuation modes of the physical degrees of freedom. To rephrase, the boundary theory comes to describe the fluctuations - which is part of the bulk theory physics - around the onshell background geometry. This way, the boundary description becomes dual to the bulk description. In other words, the former {\em is} the latter for the onshell physics. Note that there is a subtle but crucial difference between this and the standard viewpoint of AdS/CFT-type dualities in that in our view the boundary degrees of freedom are {\em not} alien to those of the bulk: they are part of the bulk theory.\footnote{For how the worldvolume theory manifests, see section 3.3 of \cite{Hatefi:2012bp}.}

The main goal of the present work at the technical level is to generalize Wald's entropy and conduct an in-depth study that aims at paradigmatically establishing a generic appearance of a Page curve \cite{Page:1993wv} from the first principles of pure - viz, without relying on AdS/CFT - gravitational physics. This undertaking inevitably involves conceptual aspects of the BHI paradox. BHI paradox is multifaceted problem. For instance, it can be formulated as a special case of the ``second law puzzle'' \cite{Kay:2022wpn}. B. S. Kay's interesting series of papers \cite{Kay:1998vv,Kay:1998cj,Kay:2012vw,Kay:2012xn,Kay:2022wpn} put forth a proposal of entropy expected to obey the second law of thermodynamics, the so-called matter-gravity entanglement entropy. It is enlightening to see that entanglement between different sets of entities is expected to lead to different behaviors, namely, a Page curve vs. the second law.

One convenient framework for studying BHI is the standard quantum field-theoretic {\em scattering} setup. However, a more inclusive framework will be useful for a wider range of tasks, and we believe that a Schrodinger setup provides a suitable conceptual framework.\footnote{In this work we use the setup only at a conceptual level. See \cite{Saini:2015dea} for a recent analysis in which BHI was studied by employing the setup.} In a Schrodinger setup a transition amplitude between arbitrary initial and final states is considered, with a scattering amplitude as a special case. With this setup one can pinpoint where the full quantum-gravitational account of BHI deviates from the semi-classical or Hawking's analysis. In brief, it is precise identification and specification of the initial (and final) vacuum state(s) that lead to the evaporating process' {\em dependence} on the initial collapse.   
Central to the specifications of the initial configurations are the boundary degrees of freedom (again, the degrees of freedom referred to here are genuine gravitational degrees of freedom but not directly those of the dual gauge theory, although they should ultimately be identified.): the very live and active degrees of freedom on the boundary are a chief component of specifying the vacuua and excited states. These boundary configurations will result in various corresponding bounce solutions, which in turn mediate the black hole formation and evaporation process.

Remarks on black hole hair will be useful. There are different types of hair. At the quantum level, large gauge transformations (LGTs) transform the Fock space and thereby introduce additional Fock states associated with the transformed solutions. This is because the transformed states should be counted as physically inequivalent to the original ones. (Another way of saying this is that the large gauge symmetry is part of the moduli.) It is this hair of the microstates that should be responsible for black hole entropy. Consideration of LGS is accompanied by adoption of a non-Dirichlet boundary condition. The boundary excitations are the microstates, and they give the same coarse-grained bulk geometry. These states are counted by black hole entropy,  which is analogous to the fact that statistical  entropy counts the microstates of gas particles whose property is specified by several global variables such as temperature and pressure. There is also hair arising from quantum corrections. In other words, a different kind of the hair arises from the feature that makes QFT different from the first-quantized framework: the Lagrangian (thus the field equations) is shifted by quantum corrections. We refer to \cite{Park:2019lbj} for more detailed discussion of different kinds of the hair. 

A word of clarification on 'generalization' is in order.
As well known there are several different ways of getting the `area law' of black hole entropy. One of them is based on Noether procedure, pioneered by Wald \cite{Wald:1993nt}. To better understand it we compare it with the traditional Euclidean action method \cite{Gibbons:1976ue}.  
The two methods are complementary: the former makes clear certain black hole physics, such as Page curves. The latter is applicable to a time-dependent solution, provided that it satisfies a Dirichlet boundary condition. It is not clear how to apply the Wald's method to a solution with time-dependence that cannot be "gauged away" by an LGT (see the body). For such a solution one should presumably follow the Euclidean action method, but with suitable boundary terms replacing the standard York-Gibbons-Hawking (YGH) term. This proposal is the highest-level generalization; it is at a conceptual level, and actually finding `suitable boundary terms' is likely to be a non-trivial task. Fortunately, however, it should be possible to understand essential physics by restricting to time-dependent solutions that are asymptotically flat. For this class of solutions, by `generalization,' we do not mean that a different formula be employed to compute the Wald's entropy. Instead, the generalization involves two facets. The first is the intermediate steps to derive the Wald's formula: we base the derivation on LGS. The second is a unification aspect. There is a sense in which the Euclidean thermal action method with a Dirichlet boundary condition can actually be understood as a special case of entanglement entropy computation - which later makes a connection with the Noether method based on LGS - with more general boundary conditions. (However, the most general case will again have to rely on the Euclidean method, as stated above.)

In the spirit of the highest-level generalization one can interpret the Wald's method as in line with the Euclidean action method. In other words the Wald's method is a particular way of evaluating the action which is nothing but the leading part of the free energy. This clarifies two things that the Wald's approach did not address: the integration range of the action and overall normalization of the Noether charge. Our contention is that what was referred to as the `interior boundary' - which is ultimately identified as the event horizon - in Wald's work \cite{Wald:1993nt} naturally arises by considering the Noether procedure as part of an action-evaluating procedure. With that, the action integral comes to range from the event horizon to infinity. Similarly, the overall normalization of the charge is automatically achieved.

The observations above unravel the essence of black hole entropy. While black hole entropy can be interpreted as thermodynamic entropy, it is not entirely clear what the `environment,' i.e., the thermal bath, is. There has also been expectation that the black hole entropy must be entanglement entropy (see, e.g., \cite{Solodukhin:2011gn}). What are the entities that are entangled? We show by filling some gaps in these views that both of these views are correct: for the former we identify the environment as the degrees of freedom outside a (stretched) horizon. For the latter it must again be the (stretched) horizon that divides the entangling bodies. We show that it is the non-Dirichlet sector boundary dynamics that provides the missing links. For both views the boundary degrees of freedom and their dynamics are the key.

\vspace{.3in}

The rest of the paper is organized as follows.
\vspace{.1in}

In section 2, we discuss our perspective on the semi-classical approach, Firewall \cite{Almheiri:2012rt}, and a Page curve. At the conceptual level our discussion is staged on a Schrodinger setup. With the setup it is straightforward to point out the critical limitation of the semi-classical approach. One can also put a finger on where the full quantum gravitational account of BHI deviates from the Hawking's analysis - which is the purpose of employing the setup to begin with. We take up several folklore statements against Firewall that we find obscure and/or debatable. Additional Firewall-supporting arguments from the boundary kinematics' viewpoint, including a typicality argument, are offered. We wrap up by discussing genericity of a Page curve, which arises from the genericity of a bounce solution. In section 3 we generalize the Wald charge to a time-dependent solution. We identify the black hole microstates as the states associated with gravitational boundary configurations. What makes this identification possible is, as previously stated, the fact that the boundary degrees of freedom are part of the gravitational degrees of freedom, i.e., genuine gravitational degrees of freedom. The identification of the boundary degrees of freedom as black hole microstates naturally leads to overhaul of several related subjects. This includes Wald's entropy charge and the roles of a large gauge transformation (LGT).\footnote{In some of the comments, replacing LGS by asymptotic symmetry would be more appropriate; we will not be concerned with this distinction.} Not unrelated to the issue of an isolated system, the standard Noether procedure is implemented with the fields satisfying Dirichlet boundary conditions. However, it is recently seen that non-Dirichlet conditions are as crucial and indispensable. To explore the in-depth physical meaning of the Wald's formula, we consider  configurations obtained by performing LGTs on a given configuration. For an asymptotically flat black hole spacetime, we discuss why the Wald's formula correctly captures the entropy. The LGS-based definition of entropy also reveals a limitation of the Wald's approach that may well be behind the discrepancy \cite{Hung:2011xb,Astaneh:2014sma,FarajiAstaneh:2014oju,FarajiAstaneh:2015fuz} between holographic and Wald's entropies. Mathematical attributes of a Lie derivative play an important role in the key analysis in section 3. The observations in the present work imply that black hole entropy is entanglement entropy in a thermodynamic setup. In section 4 we conclude with a discussion of a few points to be tied up in future works.

\section{On BHI, Firewall, and Page curve}

In this section we discuss several subjects revolving around black hole hair and entropy before carrying out a more technical analysis in the subsequent section. Although there is no shortage of discussions of these topics in the literature, we offer new insights based on the foliation-based approach of gravity quantization \cite{Park:2014tia}\cite{Park:2019amz}. We split this into three tasks.

Firstly, we zero in on the deficiency of the semi-classical analysis that led to the evaporating process’ independence of the initial collapse state. As previously noted, the deficiency lies in that the metric is not quantized, thus overlooking in particular the fact that the in- and out- vacua may not be equivalent. We corroborate the view put forth in \cite{Park:2017wiw} and \cite{Park:2019lbj} regarding how the full quantum gravitational account avoids this problem. Although we ultimately narrow down to a {\em scattering} amplitude to remedy the deficiency, for a broader perspective, it is useful to start with a more general transition amplitude. In principle, one may compute a transition amplitude between arbitrary initial and final states, which can be written in informal notation as $<final|initial>$. Here, by states are meant the Schrodinger picture states. The following connection to a Coleman-De Luccia (CdL) bounce solution (or, more precisely, its gravitational analogue\footnote{In \cite{Park:2019lbj} a simplistic gravitational bounce solution based on a shock wave was constructed, postponing construction of a more realistic solution. In fact, the construction should be possible by following \cite{Burda:2015yfa} of which we have become aware recently. \la{gbs}}) will be established below:
\bea
|false\; vac\!>=|initial\;collapse\!>\quad,\quad  |true\; vac\!>=|final\;white\; hole\!>.
\la{ftv}
\eea
With this, the amplitude of one's interest can be written as
\bea
<final|initial>=<\!true\; vac|false\; vac\!> 
               =<\!final\;white\; hole|initial\;collapse\!>. \la{fta}
\eea

The second task concerns Firewall. We debate several common statements made in the spirit of the semi-classical approach.
In the Firewall literature, avoidance of BHI paradox (and Firewall) was mostly attempted {\em within} the semi-classical description. In contrast, the approach of \cite{Park:2017wiw,Park:2018xtt,Park:2019lbj} and present work is such that within the semi-classical framework the information is indeed lost; it takes full quantum gravity to avoid BHI paradox.

The third task is to argue the relevance of a bounce solution to a Page curve. Given an initial state there should be a wide variety of final states that can be interpreted as an evaporating process. In particular, there will be a class of solutions that lead to the transition without being over-determined by the final state conditions. They should be gravitational analogues of a Coleman-De Luccia bounce solution. The boundary dynamics and LGS make the meaning and feasibility (and/or likelihood) of an isolated system sophisticated. Information can leak out, which is mediated by a bounce solution, for an un-isolated system, which should satisfy a non-Dirichlet boundary condition. To make things intertwined, a generic bounce solution should have a non-Dirichlet boundary condition. Not unrelated, the charge of an LGS-transformed system will not be conserved: the non-conservation means the current flows across the asymptotic boundary \cite{Park:2018xtt}.

\subsubsection*{Limitation of semi-classical analysis}

Although a semi-classical analysis has its own virtues, care must be exercised in assuming its range of validity. A statement commonly found in the literature is that a semiclassical description should be valid until the size of the black hole reaches a Planckian scale. As an infalling object approaches the horizon, however, its kinetic energy will become very large and Planckian physics will be involved.

To our view, the best way to see why the semi-classical approach adopted by Hawking led to BHI loss is to contrast it against a full quantum gravitational description and pinpoint where the latter deviates from the former.\footnote{Related discussions were presented in the earlier works, such as \cite{Park:2017wiw} and \cite{Park:2019lbj}. It is, however, the discussion around eqs. \rf{sauv} and \rf{sanuv} below that most clearly and explicitly ``pins the issue down."} In the semi-classical framework one quantizes the matter fields, treating the metric as a fixed background. In addition, only the vacuum associated with the Dirichlet boundary condition is considered. (At least that is what was conceived.\footnote{In a Hartle-Hawking vacuum, for instance, the boundary degrees of freedom are excited. Since a Dirichlet boundary condition was always assumed, this raises a question on internal consistency of the analysis.}) In short, it is the dynamism of the metric in the full quantum gravitational description that makes a difference: once one considers a dynamical metric, diffeomorphism symmetry - in particular, the large gauge symmetry - comes into play. Since the LGTs are nothing but the hair of the black hole, one ends up removing the hair as a result of considering fixed geometry.

The quantities computed in semi-classical analyses typically take the form of
\bea
<BH_0|(...)|BH_0> \la{mza}
\eea
where the subscript zero denotes a fixed geometry in the sense that the metric is not quantized; the ellipsis denotes an observable, or product of operators, more generally. Decay of a black hole mediated by a gravitational analogue of a CdL bounce solution is central to the proposed resolution of  BHI paradox in \cite{Park:2019lbj} employing a full quantum gravitational description. For the resolution, one should consider quantized metric and two backgrounds,  'black hole' and, say,  'Minkowski spacetime'. Decay of the former to the latter can be related, at an intuitive level, to a gravitational analogue of a CdL bounce solution. (The connection will be made more precise below.)
Let us denote the Fock vacua corresponding  to the aforementioned two backgrounds, the black hole vacuum $|BH\!>$ and Minkowski vacuum $|M\!>$, respectively. Note that quantization of the metric being considered, the subscript zero has been removed. For the scattering amplitude discussion later these vacua take the forms of in- and out- states. Naively, the false vacuum of the Coleman-de Lucci formalism corresponds to $|BH\!>$ and the true vacuum to $|M\!>$. We will explain shortly why this view is naive, and what a more proper view should be; for now we proceed with these identifications.

To be specific let us narrow down to a scattering amplitude setup.
In the present lingo, the Hawking analysis led to the paradox, roughly speaking, because the amplitude between the two different vacua, of the form
\bea
<M|(...)|BH>,
\eea
was not considered. There are several rationales for considering such decay from the black hole vacuum to the Minkowski vacuum. Firstly, the transition is all but what is implied by evaporation of the black hole. Secondly, from the statistical point of view, the original amplitude, eq. \rf{mza}, is a highly restrictive special case of measure zero in that the boundary dynamics was not taken into account. Besides these reasons, note that energy conservation does not stand in the way since it only applies to an isolated system: once one considers large gauge symmetry, non-Dirichlet sectors become relevant, and the system is no longer isolated in the standard sense of the word.

Although starting with an eternal black hole is heuristic as well as useful to motivate the connection with the CdL formalism, the identification of the true and false vacua above is naive and imprecise: the identification of $|BH\!>$ as a true vacuum is not adequate, given that the black hole state decays.\footnote{Indeed it is well known that it is Unruh vacuum that describes a decaying BH. The Unruh vacuum was obtained by carefully conducting perturbative analysis around a collapsing solution. Although it shuold be possible to consider a bounce solution associated with decay of Unruh vacuum, the analysis will enevitably be more complicated. To some degree, the simplification involved is analogous to the one achieved by examining an Unruh effect, instead of Unruh vacuum, for BH decay.} For proper identification of the initial state (false vacuum) and final state (true vacuum), non-Dirichlet boundary conditions must be taken into account. Once the boundary dynamics is taken into account, the initial state `$|BH>$` should not be taken to be an eternal state but, more naturally and generically, a collapsing state. In other words, the correct identification of the false vacuum is an initial collapsing state. Similarly, but more subtly, what $|M>$ represents should be an asymptotic white hole state, leading to eqs. \rf{ftv} and \rf{fta}.

Before discussing the statistical aspects, let us reiterate the above in more formal notation. Typically, the amplitude considered in a semi-classical analysis takes the form of
\bea
<\Psi|\f\cdots \f|\Psi> \la{sauv}
\eea
where $\f$ collectively represents the fields and $|\Psi>$ the vacuum. Disregarded here is the inequivalence between $_{out}<\Psi_{final}|$ and $|\Psi_{initial}>_{in}$. To see the information preserved and the state remain pure, one must consider a more general form of an amplitude:  
	\bea
	_{out}<\Psi_{final}|\f\cdots \f|\Psi_{initial}>_{in} \la{sanuv}
	\eea
and take into account the fact that the in- and out- vacua can be inequivalent. For a fixed initial state, one should also consider many different final $<\Psi_{final}|$ states. With the inequivalent vacua taken into account, it is expected to have an additional factor - which contains the bounce contribution - multiplying the thermal expression: after the contraction of the fields between the in- and out- vacua one will end up with $_{out}<\Psi_{final}|\Psi_{initial}>_{in}$ and this will depend on the initial state, thus preserving the information. 
	
If one examines the state $|\Psi_{initial}>_{in}$ (which is pure) near the horizon, one will be unable to distinguish the state from a mixed state. Even though the state looks mixed or thermal near the horizon, it is pure due to the entanglement of the near-horizon degrees of freedom and those radiated away.  A Page curve from a bounce solution then reflects the entanglement between the black hole and the radiation.

\subsubsection*{Essence of Firewall argument}

The Firewall argument \cite{Almheiri:2012rt} (see \cite{Braunstein:2009my} and \cite{Mathur:2009hf} for earlier related works) is a clever application of quantum principles to a General Relativistic system, capturing essential physics near the event horizon of a black hole. According to the Equivalence Principle (EP) or, to be exact, its conventional interpretation the near horizon region must be in a vacuum state for an infalling observer. In other words, locally things must look as they would in a Minkowski spacetime, with no particles around. The core of the AMPS argument \cite{Almheiri:2012rt} is that at the quantum level it should be impossible to have entanglement (which is necessary for a smooth horizon) between the black hole interior degrees of freedom and those of the late Hawking radiation.

 One of the criticisms of Firewall is that if the infalling observer did not make any measurement before falling, then the black hole state could be an eigenstate (in particular, vacuum) of the infalling observer, hence the observer will not encounter Firewall, being consistent with the EP. However, let us suppose that this person made a measurement before infalling. Then the observer will encounter a Firewall. The EP is a classical concept and it does not depend on whether or not the infalling observer makes a measurement before falling. Furthermore, to our view one does not need to complete the experiment: any serious step of the experiment will interfere with the system and the system will not be in vacuum for an infalling observer. In the Firewall context, what is meant by ``vacuum'' is ''local vacuum," as mentioned above. Imposing such an EP-based classical notion of vacuum on a potentially highly quantum field-theoretic situation seems unnatural, even a priori. The Equivalence Principle is based on a geometric description of motion of a point particle. However, it must only be in a strictly point-particle case where the motion could be understood as a geometrical effect: once quantum field-theoretic physics is involved, a strictly point-like object is artificial, since any object will have certain spread. 

In QFT the Lagrangian gets modified by quantum corrections and this shift was crucial for generation of a trans-Planckian energy. The solutions of the field equations contain the boundary modes. We refer to \cite{Nurmagambetov:2018het}\cite{Nurmagambetov:2020ann} for details. Because of this, an issue of which state is a typical state arises: given the existence of the boundary modes, their presence will be much more generic than their absence. In other words, the case of their absence - which should correspond to a Dirichlet boundary condition and smooth horizon - must be of measure zero. In the original classical background solution, which we may call the original nonperturbative state, an infalling observer does not encounter any drama as one can check by computing contraction of a geodesic and stress tensor. With the quantum corrections this changes: with the 1IP effective action one examines a quantum-corrected nonperturbative state, i.e., a solution of the quantum action. A quantum-level solution contains the $\hbar$-order boundary modes and the solution displays a trans-Planckian energy scaling.

Although the Firewall argument \cite{Almheiri:2012rt} apparently assumes the semi-classical approach, it implicitly employs quantum gravitational effects to a certain degree \cite{Nurmagambetov:2018het}. It may seem that what’s important for the trans-Planckian energy is overall quantum effects, regardless of whether they come from matter or graviton fields, but not necessarily the quantum
gravitational effects. In other words, although the AMPS argument clearly involves quantum field-theoretic ingredients, whether or not it involves quantum gravitational ingredients is subtle. Before examining this further, let us recall what's shown in \cite{Nurmagambetov:2020ann} and its earlier sequels. Strictly speaking, it is the quantum effects
(regardless of whether they are matter- or graviton- originated) plus the
metric back reaction that are important for the trans-Planckian energy. The
fact that one considers the metric back reaction reflects dynamism of the metric. Once one considers a dynamical metric and matter quantum effect,
there is no rationale to exclude the graviton loop effect, hence the relevance of
the quantum gravitational effects \cite{Nurmagambetov:2020ann}. The AMPS argument in \cite{Almheiri:2012rt} was based largely on the semi-classical framework
but still led to a trans-Planckian energy. The following must be the reason that AMPS argument indeed involves dynamism of the metric. Their argument involves a Schwarzschild observer and a Kruskal observer. In other words their setup involves a coordinate transformation. One does not consider symmetry transformation for a nondynamic field. The fact that they did implies that the metric is dynamic in their argument. In \cite{Nurmagambetov:2020ann}, there is an interplay between the metric part and matter part: the relevant zero modes are those of the metric and it was the quantum correction terms, such as $R^2$ terms, that were important for the structure of the solution. Meanwhile, for the stress tensor analysis it was the matter kinetic term that yielded the trans-Planckian energy. Presumably, this is how the AMPS Firewall argument is realized at the quantitative quantum gravitational level.

\subsubsection*{A Page-like curve from a bounce solution}

We give two independent arguments that conceptually facilitate the arrival of a Page curve, one with the help of AdS/CFT and the other based on LGS. The discussion of the second argument will be brief, with details postponed to the next section. For now we restrict to relatively simple cases. More general cases will be discussed in section 3. In both arguments, it is the event horizon area that captures the entropy at the end. These arguments clearly demonstrate genericity of a Page curve.

Consider a black hole and split the region into the black hole itself and the outside. Say one computes the entanglement entropy of the black hole after tracing out the states associated with the inside. With gauge-fixing, the physical states must have support at the boundary \cite{Park:2014tia,Park:2014noa,Park:2016zgt,Park:2018vci,Park:2019amz}, so we are effectively integrating over the boundary theory. Invoke the Ryu-Takayanagi proposal (see, e.g., \cite{Nishioka:2009un} for a review) with gradually increasing boundary area: if one integrates over the entire boundary the corresponding extremal surface is the event horizon.  In other words, initially one integrates over the boundary theory (recall that the boundary theory is nothing but the bulk theory of the physical states; see the comments in the introduction), but then one can consider the minimal surface in the original gravity theory. As long as there is a Killing vector outside of the black hole, the area of event horizon captures the entanglement entropy \cite{Park:2019lbj}. This obviously leads to a Page curve. One can also deduce a Page curve by considering an LGT, which is the subject to be studied in detail in section 3.  When the gauge parameter dies out at infinity, the gauge transformation is just redundancy. In contrast, when the parameter is "large" the gauge transformation should count the degrees of freedom.

There is a crucial issue of how generically a bounce solution represents the black hole evaporation process. Let us take a step back and consider a transition amplitude between arbitrary initial and final states. In the present case, the initial state in which we are interested is a boundary configuration that will later lead to a black hole formation. Given such an initial state, there will be a wide variety of final states that can be interpreted as an evaporating process. Some of which will be, say, more complex than the ``minimal'' process that we turn to in section 3. There should be a class of solutions that are guaranteed to lead to the transition in the sense that the initial and final conditions are not over-determined, and thus provide a more controlled setup. These should be gravitational analogues of Coleman-De Luccia bounce solutions. Because a bounce solution is time-reversal-symmetric, it is only the initial (as opposed to initial and final) conditions that one gets to impose. The solution will exist for a generic initial condition that will be of a non-Dirichlet type. Therefore, the decay will occur generically.

\section{ Wald's entropy: revisit and generalization}

The primary goal of this section is to quantitatively understand the origin of black hole entropy and come up with a definition of entropy that can be applied to an arbitrary black hole, including time-dependent ones. As well known, the area entropy law can be derived in several different ways. We focus on two of them: the first is based on a Euclidean action and the second on Wald's method. Below, Wald's entropy is generalized to one based on large gauge symmetry. Given that in the Wald's original paper \cite{Wald:1993nt} the starting point was gauge symmetry, our proposal is not entirely new. However, the crucial new ingredients are the boundary dynamics and large gauge symmetry. (See, e.g., \cite{Azeyanagi:2009wf,Liu:2021hvb,Ma:2022nwq} for works in which the importance of boundary terms in black hole entropy computation was noted.) Also, the starting point is entanglement entropy calculation, unlike the original Wald's derivation. The new definition is motivated in part by examining an LGT on a given configuration. We employ an {\em active}, as opposed to passive, point of view for convenience. The mathematical fact that a Lie derivative is a tensor type-preserving operation - a feature not shared by a
covariant derivative\footnote{This is so in the component notation. In the coordinate-free notation, tensor types are preserved both by Lie and covariant derivatives (but not by a covariant differential) \cite{Kobayashi}.} - plays a role in the manipulation.

Several cautionary remarks regarding the generalization are in order. Although large gauge variation takes the same form as ordinary gauge variation, it is only the large gauge symmetry that is relevant for the subsequent Noether charge analysis. In other words, the ordinary gauge symmetry is redundancy to be fixed, but not something to be counted as contributing to enumeration of the physical states. In contrast, LGTs enumerate the space of the inequivalent solutions connected by the transformations of a solution with a given boundary condition, such as a Dirichlet: the LGT-deformed solutions belong, unlike ordinary-gauge-transformed ones, to the moduli of the theory.

The results of the present work shed light on the long-evasive question: what is the origin of the black hole entropy and what is the identity of the responsible degrees of freedom? The boundary degrees of freedom with LGS and their dynamics naturally suggest that they should be responsible for the black hole microstates and the associated entropy.\footnote{Another way to see this is through a path integral: the measure should ultimately be over the physical states, which have support on the boundary \cite{Park:2014noa}.} The present analysis reveals the microscopic origin of the entropy - which is not revealed by the direct evaluation of the Euclidean action - as the solutions connected by LGTs and their Fock spaces.






\subsection{Generalization of Wald's entropy}

Being a Noether charge, Wald's entropy is distinct, to start with, from thermal entropy. It is an independent definition of a charge, hence the relation between the two is not a priori clear. (This issue was addressed to some extent in the original Wald's paper \cite{Wald:1993nt}.) For a stationary black hole, one can adopt the Wald's definition dependent on a Killing vector. However, it is not clear what to do with a time-dependent solution. Here we show that there exists a way of obtaining Wald's entropy that avoids these problems. This is achieved by our procedure based on entanglement entropy computation followed by LGS analysis. The establishment of the precise connection between the generalized Wald' entropy and the entropy by Euclidean action method makes clear the reason that Wald entropy formula correctly captures the black hole entropy in relatively simple (but important) cases. The key to this is 'unification': it turns out that an LGT of the metric followed by functional expansion of the action - which can be regarded as an extension of the Wald's method - provides a more or less `unified' way of obtaining the `area law'. There are several other unclear features in the Wald's approach. One of them is that the starting point is a Neumann action since the YGH term is not included. In fact, employing a Neumann action can be justified by the generalized definition, as we discuss below. Another unclear feature is the manner in which the `interior boundary' was introduced in \cite{Wald:1993nt}: ultimately, the event horizon serves as the interior boundary, but the justification of the interior boundary is not clear to us. Below we address this as well.

Let us introduce, for convenience, the terms `Dirichlet and Neumann actions'. The Dirichlet action is defined by
\bea
S_D \equiv S_{EH}+S_{YGH} \quad,\quad  S_{YGH}\equiv 2\int_{\pa {\cal V}} d^3x \sqrt{|h|}\, \ve K
\la{awgh}
\eea
where $\ve=-1$ for spacelike boundary ${\pa {\cal V}}$ and the subscript $D$ denotes the Dirichlet boundary condition. 
In 4D, the Neumann action amounts to not adding the YGH term (in general, i.e., in $D\neq 4$, some boundary terms are required \cite{Krishnan:2016tqj}):
\bea
S_N=S_{EH}. \la{neeh}
\eea   
To set the stage for the analysis below, let us also recall the following standard steps of thermodynamics that relate the thermal partition function to entropy.
As well known, the maximum value of  Gibbs entropy  $S=-k_B \Sigma_i \r_i\ln \r_i$ (as well as entanglement entropy) occurs when the density operator $\r$ takes the thermodynamic form \cite{Sakurai}. The thermal entropy follows from the partition function $Z$ through
\bea
S_{th}= -\fr{\pa F}{\pa T} 
=k_B\Big(1+T \fr{\pa }{\pa T}\Big) \log Z 
=k_B\Big(1-\b \fr{\pa }{\pa \b}\Big) \log Z  \la{tenq}
\eea 
where $F$ denotes the Helmholtz free energy
\bea
F=-k_B T\log Z.  \la{fenq}
\eea
In the Euclidean path integral, the leading term for $F$ (setting the factor $k_B T$ aside) is the action. (The area law was also shown to follow from one-loop analysis of renormalization of Newton's constant \cite{Kabat:1995eq}. See \cite{Solodukhin:2015hma} for further results and subtleties.) In the Euclidean action method one takes the Dirichlet action and directly evaluates it over a Schwarzschild solution \cite{Hawking:1995fd}\cite{Poisson}. For entanglement entropy computation of a black hole - which is, as we will see, related to the Wald's entropy, we divide the bulk into two regions, the black hole and its outside; $\r$ represents the reduced density operator obtained by tracing out the interior states.  

To recall how the area law is reproduced in the Wald's approach, let us first start with arbitrary variation:
\bea
\d S_N=\int \left(R^{\m\n}-\fr12 R g^{\m\n}\right)\d g_{\m\n}+\int g^{\m\n} \d R_{\m\n}.
\eea
Note that for the Wald's procedure and its generalization we consider the Einstein-Hilbert action without the YGH term, i.e., the Neumann action. The Dirichlet boundary condition is not suitable once one considers LGT-transformed solutions: in general the transformed solutions will not satisfy the Dirichlet boundary condition. For gauge variation $\d_\xi g_{\m\n}=\nabla_\m \xi_\n+\nabla_\n \xi_\m$ {with a `large' gauge parameter $\xi_\m$, the above takes
	\bea
	g^{\m\n} \d R_{\m\n}&=& \nabla^\m \Big[\nabla^\l(\nabla_\m\xi_\l+\nabla_\l\xi_\m)   -2 \nabla_\m \nabla_\k \xi^\k \Big].
	\eea
	Let us focus on $v_\m \equiv \nabla^\l \d g_{\m\l}-g^{\r\s}\nabla_\m \d g_{\r\s}$. By using $\nabla_\m\nabla_\k \xi^\k=\nabla_\k\nabla_\m \xi^\k-R_{\m\r}\xi^\r$, one gets
	\bea
	v_\m 
	&=&  \Big[\nabla^\l(-\nabla_\m\xi_\l+\nabla_\l\xi_\m)    +2R_{\m\r}\xi^\r \Big].
	\eea
	The volume integration can be reduced by Stokes' theorem. The first application of Stokes' theorem reduces the integration range to the time-fixed surface; the second application to the event horizon and the spherical shell at infinity:
	\bea
	\int_{3D}d\Sigma^\m \nabla^\l(\nabla_\l\xi_\m-\nabla_\m\xi_\l)
	&\sim&\int_{2D}d\Sigma^{\m\n} (\nabla_\n\xi_\m-\nabla_\m\xi_\n)
	\la{gef}
	\eea
	where $d\Sigma^\m$ and $d\Sigma^{\m\n}$ represent the 3D and 2D boundary surface, respectively. Now let us restrict to the case in which $\xi_\m$ is taken to be a Killing vector $K_\m$ along the time direction, $\xi_\m=K_\m$. Onshell, all of the terms involving the Ricci tensor vanish, leaving
	\bea
	\int_{3D}d\Sigma^\m \nabla^\l(\nabla_\l K_\m-\nabla_\m K_\l)
	&\sim&\int_{2D}d\Sigma^{\m\n} (\nabla_\n K_\m-\nabla_\m  K_\n). \la{sdfaq}
	\eea
Upon using $K^\l\nabla_\l K^\r=\k K^\r$, where $\k$ denotes the surface gravity, and employing an appropriate normalization (which we will address below), one gets the surface area of the event horizon for the Noether charge. See, e.g., \cite{Padmanabhan:2009vy} for more details. 

\vspace{.2in}
Above, the Dirichlet action is employed and the integration range of the action is the entire spacetime. Although one evaluates the action over the original Schwarzschild solution, one could also employ a solution obtained by performing a Killing transformation, and would get the same result. (This may sound trivial but will be relevant below.) The entanglement entropy computation is more involved: it is subject not only to technical challenges but also to thorny conceptual issues. The first step is to trace out the inside of the black hole,\footnote{Strictly speaking, we believe that the degrees of freedom here should include those within the stretched horizon.} which should yield the reduced density matrix. One way to handle the tracing is to employ the so-called replica trick, which is  technically challenging. However, once one seeks a maximum value of the entanglement entropy, the general result employed above \cite{Sakurai} applies here as well: the maximum value will occur when the reduced density matrix takes the thermal form. To see this, note that the variational procedure to determine the maximum entropy (in the present case, the entanglement entropy) is ultimately over the reduced density matrix. With this, one gets \rf{fenq}. In particular, the leading term of logarithm of the partition function is again the action. What's crucial is that the integration range of the action is now between the event horizon and infinity. In other words the interior of the black hole is excluded, effectively introducing an `inner boundary'.

Let us turn to the main task of generalizing the steps leading to the Wald's Noether charge for a class of solutions obtained by conducting LGTs on a given solution. To be specific we first illustrate the idea with an Einstein-Hilbert action and a Schwarzschild solution. Afterward, we will take up more general cases. Given a background $g_{\m\n}$, consider an LGT
\bea
\gt_{\m\n}(x)\equiv g_{\m\n}(x)+\d_\xi g_{\m\n}(x) \la{metvar}
\eea
where the gauge parameter $\xi^\r$ is one that is suitable for large gauge symmetry, i.e., it does not die out in the asymptotic region. The argument $x$ is explicitly displayed to stress that an active transformation is being considered. A passive transformation corresponds to $\gt_{\m\n}(x')$ where $x'$ denotes the transformed coordinates.
Let us evaluate the Neumann action over the metric $\gt_{\m\n}$ by expanding the action around the given configuration, i.e., through functional Taylor expansion: 
\bea	
S(\gt_{\m\n})=S(g_{\m\n}+\d_{\xi} g_{\m\n})=S(g_{\m\n})+\d_\xi S+\cdots . \la{fexp}
\eea

Before proceeding, let us pause and discuss the following delicate point. For a reason to be explained $\xi_\m$ will ultimately be taken to be a Killing vector, $\xi_\m=K_\m$. It may appear that one should get zero for the variational terms in the functional Taylor expansion since the original configuration is invariant and the original value of the action was zero. However, this is not the case: the original configuration is invariant under the Killing symmetry only {\em outside} the black hole. At the event horizon the Christoffel symbols become singular.\footnote{The singularity at the EH is usually referred to as a `coordinate singularity.' All this means is that it is not a curvature singularity. To our view the term coordinate singularity is misleading at the quantum level in that it gives an impression that the singularity is not physical. If, for instance, a non-smooth EH turns out to be real, which we anticipate, the singularity will be very physical, though not a curvature singularity.} The metric should be viewed as changing according to whatever is dictated by the actual form of the Killing vector. In light of the singularity, the procedure in \rf{sdfaq} can be viewed as involving certain regularization. This raises a possibility that the functional expansion procedure may be a way of extracting the event horizon contribution with certain regularization.
As for the Euclidean method, the fact that one could employ a solution obtained by performing a Killing transformation has been reminded above: the transformed and original solutions will both yield the same result, since the relevant contribution comes from the asymptotic region. It is tempting to use this as a rationale to employ a Killing-transformed solution for the entanglement entropy computation too. To a certain extent (but not entirely), this is circular logic since we want to establish that the two approaches yield the same value of the entropy. Even with this step it is still subtle to establish what we intend to establish. This is because whereas it is clear for the thermal entropy that one is computing the same entropy as the untransformed one, in the case of the entanglement computation it is not at all clear whether or not one is computing, by considering the transformed solution, the same entropy as that of the original solution. Our stance on this matter is that the Noether method is nothing but evaluation of the action over the original solution (the solution before the Killing transformation) even though it employs the Killing-transformed one. In other words, the Noether method is a way to extract the singular contribution from the EH\footnote{The relevance of the EH in entropy computation is long known in other methods, such as those of \cite{Rovelli:1996dv} and \cite{Solodukhin:1998tc}. In the present framework both the EH and asymptotic boundary play roles: LGS is a symmetry associated with the boundary, whereas the nonzero contributions come from the EH.} of the original solution. The procedure must be computing this contribution through certain suitable regularization - which is associated with \rf{sdfaq} - of the horizon singularity. Put differently, one may take the procedure as defining the regularization.

The expansion \rf{fexp} can also be written by explicitly using the fact $\d_{\xi}={\mathscr L}_\xi$ where ${\mathscr L}$ denotes a Lie derivative, as
\bea	
S(\gt_{\m\n})= e^{{\mathscr L}_\xi}S(g_{\m\n})=S(g_{\m\n})+{\mathscr L}_\xi S+\cdots . \la{fexpequi}
\eea
The Wald's formula corresponds to the linear part of $e^{{\mathscr L}_\xi}$, i.e., an infinitesimal case. In general, the leading term $S(g_{\m\n})$ will not vanish. It does vanish for a Schwarzschild solution, and does not concern us for now; we come back to this later. The $\xi$-linear term ${\mathscr L}_\xi S$ yields, as briefly reviewed above, the area entropy when $\xi$ is taken to be the Killing vector $K$. We now show that the higher-order terms vanish, and thus do not contribute to the evalaution of the action. Let us examine them, starting with $\sim {\mathscr L}_\xi^2 S$:  

\bea
{\mathscr L}_\xi^2 S=\int_{\cal V} {\mathscr L}_\xi \Big({\mathscr L}_\xi (\sqrt{-g}\;{\cal L})\Big).
\la{liesq}
\eea
For further manipulation note that a Lie derivative is a tensor type-preserving operation \cite{Kobayashi}. Applied to  eq. \rf{liesq}, this means that both $\sqrt{-g}\,{\cal L}$ and  ${\mathscr L}_\xi (\sqrt{-g}\,{\cal L})$ (as well as ${\mathscr L}_\xi^2 (\sqrt{-g}\,{\cal L})$) are scalars.\footnote{In general, this is not strictly true since $\sqrt{-g}$ is a tensor density. However, with $\nabla_\m \xi^\m=0$, the quantities under consideration do transform as scalars.} On account of this, the above can be written as 
\bea
=\int_{\cal V} \xi\cdot \nabla \Big({\mathscr L}_\xi (\sqrt{-g}\;{\cal L})\Big)
\eea
which, in light of $\nabla_\m \xi^\m=0$ \cite{Park:2018xtt} (which should be valid even for a large gauge parameter), becomes a total derivative expression with the following further manipulations: 
\bea
&&=\int_{\cal V}  \nabla^\m \Big(\xi_\m{\mathscr L}_\xi (\sqrt{-g}\;{\cal L})\Big)
=\int_{\pa{\cal V}}  d\Sigma^\m\, \xi_\m{\mathscr L}_\xi (\sqrt{-g}\;{\cal L})
=\int_{\pa{\cal V}}  d\Sigma^\m\, \xi_\m\; \xi\cdot \nabla  (\sqrt{-g}\;{\cal L})   \nn\\
&&
=\int_{\pa{\cal V}}  d\Sigma^\m\,  \nabla^\n \Big[ \xi_\m \xi_\n  (\sqrt{-g}\;{\cal L})\Big]
-\int_{\pa{\cal V}}  d\Sigma^\m\,  \xi_\n (\nabla^\n \xi_\m)  (\sqrt{-g}\;{\cal L}).   \la{som}
\eea
The first term on the second line is a total derivative (see, e.g., \cite{Poisson}), thus vanishes due to the fact that the boundary of the boundary vanishes. The second term yields a surface gravity term; because the onshell action vanishes the second term vanishes as well.

Let us pause and summarize. A priori, Wald entropy is a certain Noether charge, but not necessarily (thermal) entropy. However, it {\em appears} in the evaluation of the Euclidean action which is the leading part of the free energy. Therefore it can be said that the entropy of black hole and Wald's formula are connected to a certain degree. For a precise connection between the entropy of black hole and Wald's formula, it is necessary to more carefully examine the transformed action \rf{fexp}. In fact, this is what has just been achieved: it is only the classical action plus the leading correction, i.e., the linear term, that determines the value of the transformed action. Above, we have shown that the second-order term in \rf{fexpequi}, $\sim{\mathscr L}_\xi^2 S$, does not contribute. The rest of the higher-order terms do not contribute either, which we now show. 

This time we consider the case in which the onshell value of the action does not vanish. The higher-order terms do not contribute to the entropy either because they produce terms that vanish due to the fact boundary of boundary vanishes or are removed by the step of eq. \rf{tenq}. To see this pattern, let us examine the next order $\sim {\mathscr L}_\xi^3 S$:
\bea
{\mathscr L}_\xi^3 S =\int_{\cal V} {\mathscr L}_\xi \Big({\mathscr L}_\xi^2 (\sqrt{-g}\;{\cal L})\Big)    
=\int_{\pa{\cal V}}  d\Sigma^\m\,  \nabla^\n \Big[ \xi_\m \xi_\n  ({\mathscr L}_\xi \sqrt{-g}\;{\cal L})\Big]
-\int_{\pa{\cal V}}  d\Sigma^\m\,  \xi_\n (\nabla^\n \xi_\m)  ({\mathscr L}_\xi \sqrt{-g}\;{\cal L}).
          \nn\\ \la{liequb}
\eea
The first term on the right-hand side vanishes due to the fact that the boundary of the boundary vanishes, and the second term can be rewritten as
\bea
=-\k\int_{\pa{\cal V}}  d\Sigma^\m\,   \xi_\m  ({\mathscr L}_\xi\sqrt{-g}\;{\cal L}).
\eea
This is the same, up to $-\k$, as the right-hand side of the second equality of \rf{som}, and one can repeat the procedure. Note that unlike the pure Einstein-Hilbert case, the leading term (i.e., the onshell value of the action) may not vanish. The extra contributions due to the non-vanishing action are removed through \rf{tenq}. To see this, let us schematically write these contributions as
\bea
\b (\cdots) \la{blin}
\eea
where $(\cdots)$ denotes an expression independent of $\b$, which should be the case for an stationary configuration. Obviously, this form of the term in the entropy is removed by \rf{tenq}.

\subsection{Salient points}

As analyzed above, the Wald's method can be regarded as a particular way of evaluating the action which is nothing but the leading part of the logarithm of the partition function obtained by tracing out the black hole interior. Thus, the Wald's method is in line with the Euclidean action method, and is a special case of the highest-level (conceptual) generalization mentioned earlier. Without the integrating-out procedure, there would be no `interior boundary': the contribution would come only from the asymptotic region, wherein the integrand is regular and would yield a vanishing result. With the interior integrated out, however, the event horizon becomes relevant and the total derivative terms yield a non-vanishing contribution. 

Sometimes the BH entropy is understood as thermal entropy and other times as entanglement entropy. What the anaysis above reveals is that one needs both of these ingredients at the same time: the result in the present work leads to a view that black hole entropy is entanglement entropy in a thermodynamic setup.

Consider a collection of solutions connected to the bulk Schwarzschild geometry by LGTs. Inequivalent states notwithstanding, these solutions must be treated as one ``bundle": although these solutions have time-dependence, each of those solutions has a Killing vector. These solutions share a global property, the entropy. This is analogous to the case of gases: there are many microstates represented by several global variables, such as temperature and pressure and, of course, entropy. How does one actually calculate the global property, the entropy, of a given bundle? This can be expedited by resorting to the Schwarzschild geometry, which can be viewed as a representative of the group of solutions connected by LGTs. Then the charge, i.e., entropy, is associated with the well-known form of the Killing vector. It will of course be possible to directly show that each of those solutions has the same charge. One just needs to transform the event horizon and Killing vector according to the LGT under consideration so that all of the relevant steps in the Noether procedure are covariantly transformed.

One may wonder why one must set the gauge parameter to be a {\em Killing} vector to compute the entropy. The answer lies in the conservation requirement of a charge. Namely, to {\em define} a charge one must consider a Dirichlet boundary condition and charge conservation. One may raise that the YGH term was not present in the original derivation of Wald's entropy. Recall, however, that the definition of a charge is subject to an ambiguity: it is up to total derivative terms. We believe that a more proper setup for deriving Wald's charge should be a Dirichlet action. Once the charge is defined, one can consider the case of the parameter set to a non-Killing vector. Such a solution would describe a 'leaking' situation. To identify what is leaking one must first consider the Killing case with a Dirichlet boundary condition and define the charge. (More on this in the conclusion.)

Lastly, let us discuss an implication of our result for the discrepancy between holographic and Wald's entropies for higher-derivative gravity theories. It was observed in \cite{Hung:2011xb} and \cite{FarajiAstaneh:2015fuz} that for agreement between holographic and Wald entropies it is crucial to have ``rotational symmetry" or ``$O(2)$ symmetry". The present result is not only in line with this but also sheds light on the reason behind. Consider a higher-derivative gravity theory. It should be the total derivative terms that are a potential source of this discrepancy, since as mentioned above only singular terms would contribute. The presence of the rotational symmetry must be linked with the existence of a Killing vector. As a matter of fact this observation was made in \cite{Hung:2011xb}: ``In the holographic framework, when the entangling surface $\Sigma$ has a rotational symmetry boundary, this typically extends to a symmetry about a bulk surface $m_{\Sigma}$. The latter then naturally becomes the extremal surface in calculating the holographic EE. In such a situation, it also appears that upon analytically continuing back to the Minkowski signature, the rotational symmetry will become a Killing symmetry, but further that $m_{\Sigma}$ becomes the bifurcation of a Killing horizon in the new Minkowski
signature spacetime. That is, the resulting bulk geometry has the structure of a black hole." If there is a time-like Killing vector, the higher-derivative terms will yield contributions that will be removed by the step of \rf{tenq}. To see this, let us go to a coordinate system in which the time coordinate becomes cyclic \cite{Misner}. The contribution to the logarithm of the partition function will take the form of \rf{blin} and thus is removed by the procedure of \rf{tenq}. If the time-dependence of the solution is complex enough to prevent a Killing symmetry, this will not be the case, and extra contributions that go beyond Wald's entropy are expected.\footnote{Although the original Wald's definition requires the presence of a Killing vector, the charge was also given in subsequent works by an expression based on a functional derivative with respect to the Riemann tensor. The present result implies that a formal entropy calculation based on such a definition will not reproduce the entropy based on an Euclidean action.}

\section{Conclusion}

The boundary theory is nothing but the theory of the gravitational boundary degrees of freedom. Given this, the entropy of the boundary theory must naturally be equal to the entropy of the bulk theory. After presenting our perspectives on various matters in section 2, we have examined several aspects of black hole entropy calculation and observed that the Wald's method is closely tied with free energy computation. Although one usually considers Wald's Noether charge entropy in the holographic context, the analysis in the present work raises a possibility that it may not always be the same as the entropy based on the Euclidean action. In our view, it is the entropy based on the Euclidean action that is more fundamental. 
\vspace{.2in}

More work is required to tighten several loose ends:
\vspace{.1in}

In this work we have focused on two specific boundary conditions: one is a Dirichlet-type and the other a Neumann. While it should certainly be a worthy effort to try to come up with the most general form of a boundary condition, it will likely be a difficult task. The present results suggest a simpler and more practical approach. Given a boundary condition, one should presumably consider all other boundary conditions connected by LGTs; one may consider each solution with LGTs as one bundle. For the most thorough treatment one will have to consider transitions between such different sectors. However, it should be possible to understand a large chunk of the essential physics by focusing on a few sectors.

The questions of what kinds of different charges LGTs lead to and their dynamics are also worth further study. Given that a Dirichlet boundary condition is required to define a charge, only time-independent LGTs can be used to yield different charges. Those LGT-transformed solutions will have the same entropy as well as additional charges at the same time. In a non-conservation situation, it may well be both entropy and the other charges that may leak. For entropy leaking, it will be the horizon area that provides a probe. It will be of interest to explore and conduct a quantitative analysis of these aspects of the boundary dynamics.

\newpage

\end{document}